\documentclass[12pt]{article}
\usepackage{amsmath, amssymb, slashed, epsf, color, graphicx,cite}
\usepackage{epsfig}

\def\met{E_T \hspace*{-1.1em}/\hspace*{0.5em}}
\def\gl{\tilde{g}}
\def\neu{\tilde{\chi}_1^0}

\begin{document}

%%%%%%%%%%%%%%%%%%%%%
%%%% Title page %%%%%
%%%%%%%%%%%%%%%%%%%%%
\begin{titlepage}
\begin{center}

\hfill IPMU14-0071 \\
\hfill KEK-TH-1718  \\

\vspace{1.5cm}
{\large\bf Compressed SUSY search at the 13 TeV LHC using kinematic correlations and structure of ISR jets}

\vspace{1.5cm}
{\bf Satyanarayan Mukhopadhyay}$^{(a)}$,
{\bf Mihoko M. Nojiri}$^{(a,b)}$,\\ 
{\bf and}
{\bf Tsutomu T. Yanagida}$^{(a)}$\\

\vspace{1.0cm}
{\it
$^{(a)}${Kavli IPMU (WPI), The University of Tokyo, Kashiwa, Chiba 277-8583, Japan} \\
$^{(b)}${KEK Theory Center and Sokendai, Tsukuba, Ibaraki 305-0801, Japan}} \\[2cm]

\abstract{The LHC search for nearly degenerate gluinos and neutralinos, which can occur, for example, in SUSY axion models, is limited by the reduced missing transverse momentum and effective mass in the events. We propose the use of kinematic correlations between jets coming from initial state radiation (ISR) in gluino pair production events at the 13 TeV LHC. A significant improvement in the signal to background ratio is obtained for the highly compressed gluino-neutralino search, by using cuts on the rapidity and azimuthal angle separation between the pair of tagged jets with the highest transverse momenta. Furthermore, the distribution of the azimuthal angle difference between the tagged jets in the $\gl \gl + $jets process is found to be distinctly different from the dominant background process of $Z+$jets. We also find quark and gluon jet tagging methods to be useful in separating the signal, which contains a higher fraction of gluon initiated jets compared to the dominant backgrounds.}

\end{center}
\end{titlepage}
\setcounter{footnote}{0}

%%%%%%%%%%%%%%%%%%%%%%%%
%%%%% Introduction %%%%%
%%%%%%%%%%%%%%%%%%%%%%%%
\section{Introduction}
The search for supersymmetry (SUSY) is one of the primary focus of the Large Hadron Collider (LHC) experiment's endeavour to find physics beyond the standard model (SM). Production of SUSY particles has been looked for at the previous runs of the LHC in several possible final states involving highly energetic jets and large missing transverse momentum ($\met$), multiple leptons or disappearing charged tracks. No significant deviation from the SM predictions has been found so far, and lower limits on coloured SUSY particle masses have reached the TeV scale in many SUSY breaking scenarios from the 8 TeV LHC search. Therefore, possible ways in which the conventional searches can miss SUSY particles are now being explored in detail. One such possibility is the so called compressed spectrum, in which the mass difference between the initially produced strongly interacting SUSY particle (squarks and gluinos) and the lightest SUSY particle (LSP) obtained at the end of a decay chain (neutralinos) is small, thereby leading to lower values of $\met$. Such a compressed spectrum is predicted in certain SUSY breaking models, for example, SUSY axion models~\cite{Nakayama} and SUSY broken geometrically in extra-dimensions~\cite{Murayama-compact}. In the wake of the discovery of a 125 GeV Higgs-like boson at the LHC, SUSY scenarios where only the gauginos are light and the sfermions are much heavier~\cite{Wells} have received a lot of attention. These scenarios, dubbed as pure gravity mediation~\cite{PGM} or mini-split SUSY~\cite{mini-split} have a gaugino spectrum as in anomaly mediated SUSY breaking (AMSB)~\cite{anomaly}, with the masses of gluino, wino and bino being proportional to the corresponding 1-loop beta functions of the gauge couplings. However, as pointed out in Ref.~\cite{Nakayama}, such gaugino mass relations can get modified in SUSY axion models, in which there is an additional sizeable contribution to the gaugino masses from the F-term of the axion supermultiplet, which, when combined with the usual anomaly mediation contribution, can lead to a spectrum where the gluino and the wino LSP are nearly degenerate in mass.

An often adopted methodology to search for such a compressed spectrum is to rely on the emission of a hard jet coming from initial state radiation (ISR), and thereby enhancing the $\met$ in the event~\cite{Alwall,Han,Martin,Dreiner,Biplob,Wacker,Wells2}. Even on inclusion of such radiation, the constraints on the compressed scenarios are generically weaker. For example, in a scenario with the gluino and the neutralino as the lighter SUSY particles, and  squarks much heavier, the current bound on gluino mass is around 600 GeV in the limit of extreme degeneracy with the lightest neutralino~\cite{ATLAS,CMS}. It is therefore important to explore avenues in which the search for such compressed SUSY particles can be improved by employing more specific topology-based criterion, whereby different possible kinematic correlations between the ISR's are fully utilized. In this study we illustrate a few of these possibilities, taking the example of a highly compressed gluino-neutralino spectrum (with a mass difference of the order of 20 GeV or less).  We focus on events with the so called vector boson fusion (VBF) topology and demand at least two hard jets widely separated in rapidity (the two highest transverse momentum ($p_T$) jets are henceforth referred to as the tagged jets).  The rapidity separation between the tagged jets is found to be a useful variable in enhancing the signal to background ratio. In addition, we find that the azimuthal angle difference between the tagged jets ($\Delta \phi_{j_1 j_2}$) has a distinctly different shape in the $\gl \gl +\geq 2-$jets process, as compared to the dominant background of $Z+\geq 2-$jets. Thus  $\Delta \phi_{j_1 j_2}$ can not only be used as a discriminating variable to boost the discovery (or exclusion) reach, in the aftermath of an actual discovery, it can be used to study the spin and CP properties of the centrally produced gluinos. Finally, we also study the possible impact of discriminating quark jets from gluon jets by using the number of charged tracks and the width (girth) of a jet as variables within a Boosted Decision Tree (BDT) algorithm~\cite{Schwartz,TMVA}. The ISR jets in the signal process are found to have a larger fraction of gluons compared to the main background of $Z+$jets, the latter containing a much larger quark-jet fraction in the hardest emission.

The remaining sections of the paper are organized as follows. In Sec. 2, we briefly review a model for obtaining a compressed gaugino spectrum following Ref.~\cite{Nakayama}. In Sec. 3 we describe our analysis framework, and the details of the signal and SM background processes studied. Sec. 4 is devoted to our central results, the discussion of the kinematic correlations between the ISR jets and aspects of using quark and gluon jet tagging methods. We summarize our findings in Sec. 5.

\section{A model for compressed gaugino spectrum}
As discussed in the introduction, after the discovery of a Higgs-like boson at around 125 GeV, a lot of attention in SUSY model building has been focussed on scenarios in which the scalar superpartners  obtain SUSY breaking masses of the order of the gravitino mass, $m_{3/2}$, due to supergravity effects~\cite{PGM,mini-split} (the Higgsino mass parameter $\mu$ is $\mathcal{O}(m_{3/2})$ as well). In order to obtain a Higgs mass of the order of 125 GeV, it is then favourable to choose $m_{3/2}$ to be $\mathcal{O}(100-1000)$ TeV~\cite{Wells,PGM,mini-split,Hmass}. The gaugino masses are generated by the anomaly mediation effect~\cite{anomaly}, and are determined by the 1-loop beta functions of the gauge couplings as
\begin{equation}
M_a = \frac{-b_a g_a^2}{16 \pi^2}m_{3/2},
\label{anomaly}
\end{equation}
where, $a=1,2,3$ correspond to the $U(1)_Y$, $SU(2)_L$ and $SU(3)_C$ gauge groups and $g_a$ and $b_a$ are the corresponding gauge couplings and one-loop renormalization group co-efficients ($b_a=-33/5,-1,3$) respectively. 
Eq.~\ref{anomaly} leads to a particular hierarchy of the gaugino masses: $|M_2|<|M_1|<|M_3|$, where $M_2,M_1$ and $M_3$ denote the wino, bino and gluino masses respectively. 

In this section, following Ref.~\cite{Nakayama}, we briefly review the deformation of the anomaly mediation spectrum for the gauginos in the presence of an axion supermultiplet, which is coupled to the gauge field strength to solve the strong CP problem. In a SUSY axion model, the gauge-singlet axion multiplet $\Phi$ can obtain, from SUSY-breaking effects, an F-term $F^{\Phi}=-m_{3/2} v \epsilon$, where $v$ is the vacuum expectation value of $|\Phi|$ and $\epsilon$ is an $\mathcal{O}(1)$ constant which depends upon the details of the model. This gives rise to a contribution to the gaugino masses which is of the same order as the anomaly mediation contribution. Therefore, the gaugino mass relations implied by the AMSB effect (Eq.~\ref{anomaly}) can be modified as follows
\begin{equation}
M_a = \frac{g_a^2}{16 \pi^2}(-b_a+\mathcal{C}_a \epsilon)m_{3/2},
\label{axion}
\end{equation}
where, the $\mathcal{C}_a$'s are model dependent constants. For example, in order to maintain successful gauge-coupling unification (e.g., in an $E6$ grand unified theory), if $N_5$ pairs of Peccei-Quinn (PQ) quarks are introduced in the $5$ and $\bar{5}$ representations of $SU(5)$, then $\mathcal{C}_a=N_5$ for all $a$.  As shown in Ref.~\cite{Nakayama}, for certain values of $N_5 \epsilon \sim 2$, one finds that the wino and gluino are nearly mass degenerate, while the bino continues to be heavier than the wino. Such a value of $N_5 \epsilon$ can be achieved if we introduce 3 pairs of PQ quarks, and the multiplet $\Phi$ is a combination of two Higgs multiplets $P(+1)$ and $Q(-3)$, where the numbers in parentheses are the PQ charges~\cite{Nakayama}. In this case, $\epsilon=2/3$ and $N_5=3$, leading to a positive value for $N_5 \epsilon=2$, which then gives rise to an almost degenerate gluino and wino mass. In such a scenario only the  gluino, the lighter charged wino, and the wino-like LSP will have the best prospects of being observed at the LHC. While the very small mass difference between the charged and the neutral wino~\cite{Shigeki1} can lead to a disappearing track signature, the LHC reach in this channel is rather limited, and we therefore focus on the gluino pair production search. It is interesting to note that the wino dark matter in such a scenario can satisfy the relic-abundance requirement via the gluino-wino co-annihilation process~\cite{Strumia, Shigeki2}.
 
\section{Analysis framework}
We consider a simplified SUSY scenario with only the gluino ($\gl$) and the lightest neutralino ($\neu$) accessible at the LHC energies, the remaining SUSY particles being much heavier. In particular, we are interested in an example spectrum with extreme degeneracy between the $\gl$ and the $\neu$, and for simplicity, fix their mass difference to be $M_{\gl}-M_{\neu}=20$ GeV. The methods discussed in this paper will be of general validity in a compressed gaugino search, and the above mass splitting is chosen for illustration only. The decay mode of the gluino considered is via off-shell squarks of the first two generations to a light quark pair and the LSP, namely, $\gl \rightarrow q \bar{q} \neu$. In the absence of energetic ISR jets, since the gluinos themselves will be dominantly produced near the kinematic threshold, the jets coming from gluino decay will have very low $p_T$, and thus, most often being below the trigger threshold of the LHC detectors. As mentioned in the introduction, we focus on final states with at least two ISR jets in a VBF-type topology, i.e., on the signal  process $\gl \gl +\geq 2-$jets, where the hardest jets are widely separated in rapidity and the gluinos are centrally produced. For our numerical analysis, we choose the following two representative points above the current LHC exclusions:
\begin{itemize}
\item {\bf Point-A:} $M_{\gl}=800$ GeV, $M_{\neu}=780$ GeV
\item {\bf Point-B:} $M_{\gl}=1000$ GeV, $M_{\neu}=980$ GeV
\end{itemize} 

The SUSY mass spectra at the electroweak scale are obtained with the spectrum generator {\tt SuSpect}~\cite{suspect}. The parton-level events for the 13 TeV LHC are generated with {\tt MadGraph5}~\cite{MG5}, which are then passed onto {\tt PYTHIA6}~\cite{Pythia} for parton-showering, hadronization and decays (with the {\tt Z2} tune in {\tt PYTHIA6}~\cite{Field}). The default MLM matching algorithm~\cite{MLM} for combining the matrix-element (ME) and parton-shower jets as implemented in {\tt MadGraph5} has been used. We use the {\tt CTEQ6L1}~\cite{Cteq} parton distribution functions from the {\tt LHAPDF}~\cite{LHAPDF} library, and the factorization and renormalization scales are kept at the default event-by-event choice of {\tt MadGraph5}. For simulating the detector effects, we use {\tt Delphes2}~\cite{Delphes}, where the jet clustering is performed with {\tt FastJet3}~\cite{Fastjet}. Jets have been formed using the anti-$k_T$ clustering algorithm~\cite{Fastjet,antikt} with radius $R=0.4$. Some of the variables used for studying quark and gluon jet tagging (as discussed in Sec.~\ref{sec:qg}) have been implemented by us in the {\tt Delphes2} framework.

The dominant SM background in the jets$+\met$ channel (with no isolated charged lepton), with the number of jets $n_j \geq 2$, comes from $Z+$jets production, followed by $Z \rightarrow \nu \bar{\nu}$. The sub-dominant backgrounds include $W+$ jets, with $W \rightarrow \ell \nu$ (if the lepton is missed, mostly when its pseudorapidity is outside the tracker or muon chamber coverage, i.e., $|\eta_\ell|\gtrsim 2.5$, or the $W$ boson decays to a hadronically decaying tau lepton), and $t \bar{t}+$ jets. As demonstrated in Ref.~\cite{QCD-Backg}, the QCD background can be eliminated by using a strong cut on the $\met$ variable (we shall eventually demand $\met>300$ GeV), and by ensuring that the $\overrightarrow{\met}$ vector is azimuthally separated from the jet directions. The simulation framework used for the SM backgrounds is the same as for the SUSY signal described above. In order to obtain a sufficient number of Monte-Carlo (MC) events in the kinematic regime of our interest, we generate our event samples after strong cuts on the $p_T$'s of the two leading jets at the matrix-element level. For the dominant as well as very large $Z+$jets background, we apply  an additional generation level cut on the $\met$ variable. This makes it difficult for us to normalize our total matched cross-sections to next-to-leading order (NLO) in QCD results, since a) it requires a fully differential NLO simulation to obtain the proper K-factors after the jet-$p_T$ and $\met$ cuts and b) we found that the $\gl \gl + \geq 2$-jets matched cross-section is quite sensitive to the choice of the matching scale. Therefore, although the NLO K-factor for $\gl \gl$ production is significantly larger (around 1.9) than the corresponding K-factor for weak boson ($W,Z$) production (around 1.2), and including such a K-factor can enhance the LHC reach for gluino mass in our study, we abstain from adopting a normalization by such K-factors for the above two reasons.

Since one of the main focus of our study is the kinematic correlation between the ISR jets, and we do not use any veto on the third or higher number of parton emissions, we have carefully considered the effects of a third hard radiation by including the $\gl \gl +1,2,3-$jets ME's in our signal simulation as well the $Z+1,2,3-$jets ME's for the dominant background simulation. This takes into account any possible modification in the dijet kinematic correlations due to additional hard ISR's. For the sub-dominant backgrounds of $W+$jets and $t \bar{t}+$jets we include ME's with upto two additional partons.

\section{Results}

Having described our simulation framework in the previous section, we now discuss the different selection criteria employed to separate the gluino signal from the large SM backgrounds. We first make a preselection of events based on the following cuts:
{\bf Cut-1:}
\begin{enumerate}
\item Number of jets: $n_j \geq 2$ with $p_T^{j_1}\geq 100$ GeV and $p_T^{j_2} \geq 50$ GeV. For all other jets we demand $p_T^j \geq 20$ GeV. The rapidity coverage of the jets is determined by ATLAS calorimeter design, where the forward calorimeter covers the pseudorapidity range of $|\eta|<4.9$, as implemented in {\tt Delphes2}. However, the tracker covers only upto $|\eta|<2.5$, and therefore it is not possible to obtain the information on the number of charged tracks inside jets in the forward region. 

\item No isolated lepton (electron or muon) with $p_T>10$ GeV, within $|\Delta \eta|<2.5$. 

\item Missing transverse momentum in the event $\met > 100$ GeV.
\end{enumerate}
The jet $p_T$ cuts are applied on all processes at the ME level, and in addition the $\met$ cut is also applied while generating the $Z(\rightarrow \nu \bar{\nu})+$jets events. 

In Table~\ref{xsec} we show the cross-sections for $\gl \gl +\geq 2-$jets in Point-A and Point-B, and  for the different SM background processes after various cuts (all cross-sections are in fb units). The total SM cross-section is also shown for reference. In addition, in the column $S_{800}/B$ we show the $S/B$ ratio (where S is the number of signal events, and B is the total number of background events) for Point-A with $M_{\gl}=800$ GeV.  
%%%%%%%%%%%%%%%%%%%%%%%%%%%%%%%%%%%%%%%%%%%%%%%%%%%%%%%%%%%%%%%%%%%%%%%%%%%%%%%%%
\begin{table}[htb!]
\centering
\begin{tabular}{|c|c c c| c|c c |c|c|}
\hline

Cuts    & $Z+$jets & $W+$jets &$t \bar{t}+$jets&Total SM &\multicolumn{2}{c|}{$M_{\tilde{g}}$(GeV)}    &$S_{800}/B$\\
            &                           &                           &                &               & 800 & 1000 &    \\
\hline
\hline
Cut-1                                       &34010       &37883.8   &16035.1   &87928.90   &276.75       	 &58.85        &0.003\\
$p_T^{j_1}\geq 200$ GeV       &11923.5    &12776.3   &4142.68   &28842.48   &165.83       &35.74         &0.006	   \\
$\met > 300$ GeV                   &1880.85    &979.41   &377.15     &3237.41    &112.53      &24.85         &0.035	  \\
\hline
\hline
{\bf Cut-A} &&&&&&&\\

$M_{\rm{eff}}>1000$ GeV          &729.89    & 460.29    &217.80          & 1407.98     &71.48        & 16.06   &0.05    	 \\
$+|\Delta\eta_{j_1 j_2}|>3.5$     &23.99        & 12.72           & 2.86            & 39.57    &5.23         & 1.03     &0.13        \\
$+|\Delta\phi_{j_1 j_2}|<\pi/2$   &10.01        & 5.23            & 1.63            & 16.87     &3.07        & 0.61     &0.18   	\\
\hline
\hline
{\bf Cut-B} &&&&&&&\\

$M_{\rm{eff}}>1250$ GeV        &310.82     & 207.12        &105.90             &623.84      &42.80      &9.84     &0.07       \\
$+|\Delta\eta_{j_1 j_2}|>3.5$      &7.55        & 4.12             &1.19               &12.86        &2.55      & 0.51    &0.20      \\
$+|\Delta\phi_{j_1 j_2}|<\pi/2$    &2.91        &  1.57             &0.61                &5.09          &1.44      & 0.29    &0.28      \\
\hline
\hline
{\bf Cut-C} &&&&&&&\\
$M_{\rm{eff}}>1500$ GeV       &138.81     & 94.62        &49.59               &283.02        &24.71       &5.87   &0.09     \\
$+|\Delta\eta_{j_1 j_2}|>3.5$     &2.61        &  1.35          &0.37                &4.33          &1.28         & 0.26  &0.30      \\
$+|\Delta\phi_{j_1 j_2}|<\pi/2$   &1.11         &   0.50              &0.16                     &1.77           &0.73        & 0.15   &0.41      \\
\hline
\hline
{\bf Cut-D} &&&&&&&\\
$M_{\rm{eff}}>1750$ GeV       &64.79       & 44.82        &22.86              &132.47        &14.08        &3.42    &0.11       \\
$+|\Delta\eta_{j_1 j_2}|>3.5$      &0.96        &   0.57             &0.20                      &1.73          &0.53         & 0.14   &0.31       \\
$+|\Delta\phi_{j_1 j_2}|<\pi/2$    &0.44        &     0.25           &0.09                      &0.78          &0.32         & 0.08    &0.41      \\
\hline
\end{tabular}
\caption{\small \sl Signal (Point-A and Point-B) and SM background cross-sections after various cuts at $\sqrt{s}=13$ TeV LHC. Cut-1 is defined above. All cross-sections are in fb units. The last column ($S_{800}/B$) shows the ratio of the signal cross-section to the total SM background cross-section for the parameter point $\{M_{\tilde{g}}, M_{\tilde{\chi}_1^0}\}=\{800,780\}$ GeV. }
\label{xsec}
\end{table}

To start with, we found it necessary to increase the $p_T$ threshold for the hardest jet to $200$ GeV and the minimum value of $\met$ to $300$ GeV, to achieve a minimal control over the huge backgrounds. After that, we show four combinations of possible choices for the cuts, Cut-A to Cut-D. The only difference in these four choices is the effective mass cut used, defined as
\begin{equation}
M_{\rm eff} = \sum_{j} p_T^j + \met,
\end{equation}
where the sum goes over all the reconstructed jets. 

\subsection{Rapidity separation between the tagged jets}
In addition to the $M_{\rm eff}$ cut, we have required the two hardest jets to reside in opposite hemispheres of the detector with a large separation in rapidity:
\begin{equation}
\eta_{j_1} \times \eta_{j_2} < 0, ~~~~~~|\Delta\eta_{j_1 j_2}|>3.5.
\label{eta}
\end{equation}
\begin{figure}[htb!]
\begin{center}
\resizebox{11cm}{!}{\input{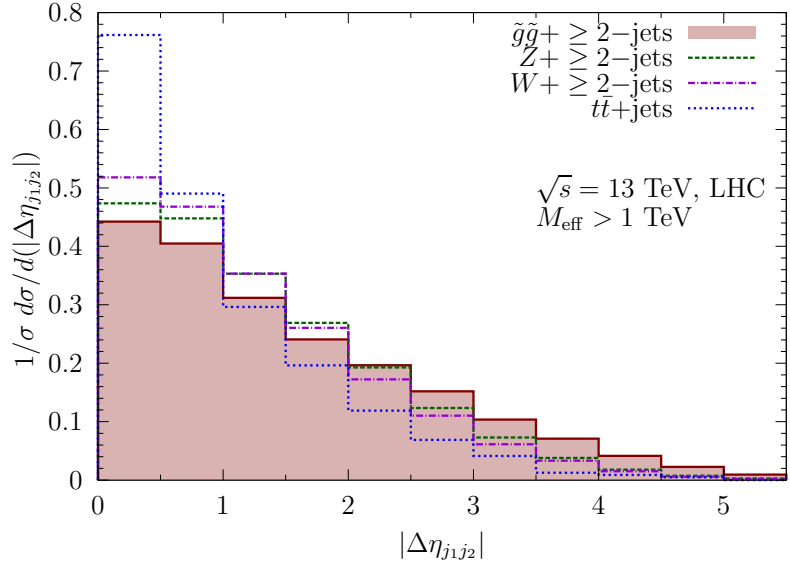}} 
\caption{\small \sl Normalized $|\Delta\eta_{j_1 j_2}|$ distributions for $\gl \gl + \geq 2-$jets in signal Point-A (shaded region), $Z+\geq 2-$jets (green dashed), $W+\geq 2-$jets (violet dot-dashed) and $t \bar{t}+$jets (blue dotted) for the $13$ TeV LHC. The distributions are shown after the jet-$p_T$, $\met$ and $M_{\rm eff}>1$ TeV cuts.}
\label{fig:deleta}
\end{center}
\end{figure}
In Fig.~\ref{fig:deleta} we show the normalized distribution of $|\Delta\eta_{j_1 j_2}|$ for signal Point-A and the SM backgrounds, after the  jet $p_T,\met$ and the $M_{\rm eff}>1$ TeV cuts. The distribution for signal Point-B has a shape very similar to Point-A.  We can clearly see from this figure that the requirement of a large rapidity separation helps to reduce the remaining $t \bar{t}+$jets background considerably, which has a higher jet multiplicity due to the presence of two b-quarks coming from top decay. Furthermore, although at least one of the tops in $t \bar{t}+$jets events has to decay semi-leptonically in order to obtain $\met>300$ GeV (either with the lepton missed, or with a hadronically decaying tau), the other top can decay in the fully hadronic mode, thereby increasing the jet multiplicity further. Therefore, we find a large fraction of events with the two hardest jets close in rapidity and the $|\Delta\eta_{j_1 j_2}|$ distribution falls off very sharply by $|\Delta\eta_{j_1 j_2}| \sim 2$.  The shape of the $|\Delta\eta_{j_1 j_2}|$ distribution for $Z+$jets and $W+$jets are very similar, and both of them have a slightly steeper fall off compared to the $\gl \gl+$jets signal. This is because the signal receives a large contribution from t-channel gluon fusion diagrams, which lead to a VBF-like topology and hence give rise to tagged jets with a large rapidity separation. We find that even though in Fig.~\ref{fig:deleta} the signal starts to show a relative excess over the background from $|\Delta\eta_{j_1 j_2}| \sim 2.5$, the choice $|\Delta\eta_{j_1 j_2}|>3.5$ gives us the best $S/B$ ratio as well as a higher reach in gluino mass. The improvement in the $S/B$ ratio with this cut is by a factor of $2.5 - 3$ across Cuts A-D, making it crucial for our search. Essentially, the cut on $\Delta\eta_{j_1 j_2}$ acts as a replacement for higher $\met$ or $M_{\rm eff}$ cuts employed in other studies~\cite{Biplob,Wacker}. Since it is difficult to obtain higher values of $\met$ or $M_{\rm eff}$ in compressed scenarios, we find the $\Delta\eta_{j_1 j_2}$ cut tailored to the signal topology considered by us.
We note in passing that we have assured the generation of MC events for all the SM backgrounds and SUSY signal processes with a reasonable statistics (corresponding to more than $100 {~\rm fb}^{-1}$ of integrated luminosity at the 13 TeV LHC), in order to minimize statistical fluctuations in the predicted cross-sections, especially after strong kinematic cuts. 

\subsection{Azimuthal angle difference between the tagged jets}
After imposing the requirement of the large rapidity separation between the tagged jets, the t-channel gluon fusion diagrams will dominate the total signal cross-section. For a given set of initial state partons (quark/gluon) and final state gluino helicities the amplitudes corresponding to different intermediate gluon helicities then give rise to interference terms which lead to specific azimuthal angle correlations between the tagged jets. In the limit of on-shell intermediate gluons this correlation is determined  by the phases of the splitting amplitudes for producing the tagged ISR jets. Since only specific combinations of intermediate gluon helicities are allowed for given final state angular momentum and CP properties, the azimuthal correlation of tagged jets is often found useful in the determination of the spin and CP properties of new particles centrally produced with two tagged jets in a VBF-like configuration, without requiring the reconstruction of the particle's decay products. For details on azimuthal correlations in Higgs and new particle production we refer the reader to Refs.~\cite{Hagiwara1,spinCP},  and to Ref.~\cite{Hagiwara2} for correlations in the QCD production of heavy quark pairs (top or bottom) in association with two jets. 

\begin{figure}[htb!]
\begin{center}
\resizebox{11cm}{!}{\input{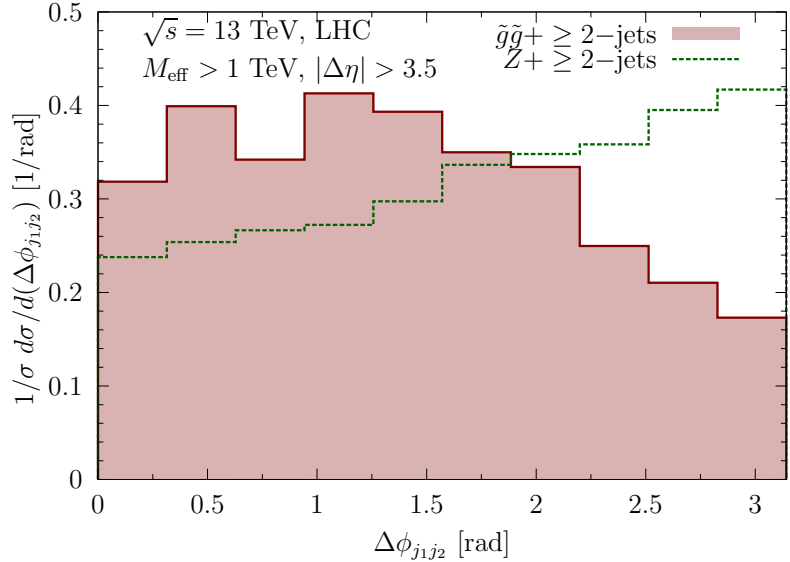}} 
\caption{\small \sl Normalized $|\Delta\phi_{j_1 j_2}|$ distributions for $\gl \gl + \geq 2-$jets in signal Point-A (shaded region) and the dominant $Z+\geq 2-$jets background (green dashed) for the $13$ TeV LHC. The distributions are shown after the jet-$p_T$, $\met$, $M_{\rm eff}>1$ TeV and $|\Delta \eta_{j_1 j_2}|>3.5$ cuts.}
\label{fig:delphi}
\end{center}
\end{figure}

In Fig.~\ref{fig:delphi} we show the distribution of the azimuthal angle difference between the two tagged jets ($\Delta\phi_{j_1 j_2}$) for the gluino signal in Point-A and the major SM background of $Z+$jets. The distribution is shown after all the basic jet $p_T,\met, M_{\rm eff}>1$ TeV cuts and the requirements on $\eta_{j_1}$ and $\eta_{j_2}$ as given in Eqn.~\ref{eta}. For $Z+\geq 2-$jets we observe a drop near $\Delta\phi_{j_1 j_2}=0$, from where it very slowly rises to $\Delta\phi_{j_1 j_2}=\pi$. On the otherhand, for $\gl \gl + \geq 2-$jets the distribution peaks at $\Delta\phi_{j_1 j_2} \sim \pi/2$, while we observe a trough near $\Delta\phi_{j_1 j_2}=\pi$. This is similar to the correlation observed for a spin-0 CP-odd particle production. Even after the cuts imposed by us, the $\gl$ pair is dominantly produced near the kinematic threshold with a symmetric colour structure, which is then an S-wave CP odd state. If indeed an excess over the SM backgrounds is observed in the search channel considered by us, it will then be of great interest to study the $\Delta\phi_{j_1 j_2}$ distribution after the $\Delta\eta_{j_1 j_2}$ cut, thereby obtaining the spin information of the produced gluino pair. Since the $\Delta\phi_{j_1 j_2}$ distributions for $\gl \gl+$jets and $Z+$jets cross at around $\pi/2$, imposing the following requirement
\begin{equation}
\Delta\phi_{j_1 j_2} < \pi/2
\end{equation}
also helps improve the $S/B$ ratio by another factor of $1.4$ in all the categories of Cuts A-D.  Thus this particular variable is beneficial for both extracting the signal as well as for making future measurement of quantum numbers.

\subsection{Jet structure: quark vs gluon initiated jets}
\label{sec:qg}
In this subsection, we explore a different search strategy for compressed gauginos, namely the use of quark and gluon jet tagging methods, to exploit the fact that the $\gl \gl+$jets signal events have a larger fraction of gluon jets compared to the main background of $Z+$jets. In particular, the hardest emission in $Z+$jets events is dominantly a quark jet. For this purpose, we have explored two variables which can discriminate gluon and quark-initiated jets,  namely, the number of charged tracks inside a jet ($N_{\rm Tracks}$), and the jet width ($w_j$). For a detailed discussion on these and several other quark/gluon tagging methods we refer the reader to Ref.~\cite{Schwartz}. Jet width (also known as girth) is defined as
\begin{equation}
 w=\sum_{i\in{\rm Jet}} \frac{p_T^i \Delta r_i} {p_T^{\rm Jet}},
\end{equation}
where, $p_T^i$ denote the transverse momenta of the jet constituent particles, and $\Delta r_i$ is their separation from the jet axis in the rapidity-azimuthal angle plane. In general, because of a larger colour factor in the splitting amplitudes, gluon-initiated jets tend to radiate more and in a bigger cone, thereby having a larger width compared to quark-initiated jets. As emphasized in Refs.~\cite{Schwartz}, the discrimination of quark and gluon jets is best achieved by combining two different types of variables: a discrete one like the number of charged tracks within the jet cone, and a continuous one like the jet width defined above. Furthermore, since the boundary between the signal region and the background region in the $N_{\rm Tracks}-w_j$ plane is non-linear, it is beneficial to adopt a multivariate analysis (MVA) strategy which can give us an optimized discriminant. For this purpose, we have employed a Boosted Decision Tree (BDT) algorithm with the help of the {\tt TMVA-Toolkit}~\cite{TMVA} in the {\tt ROOT} framework.

\begin{figure}[htb!]
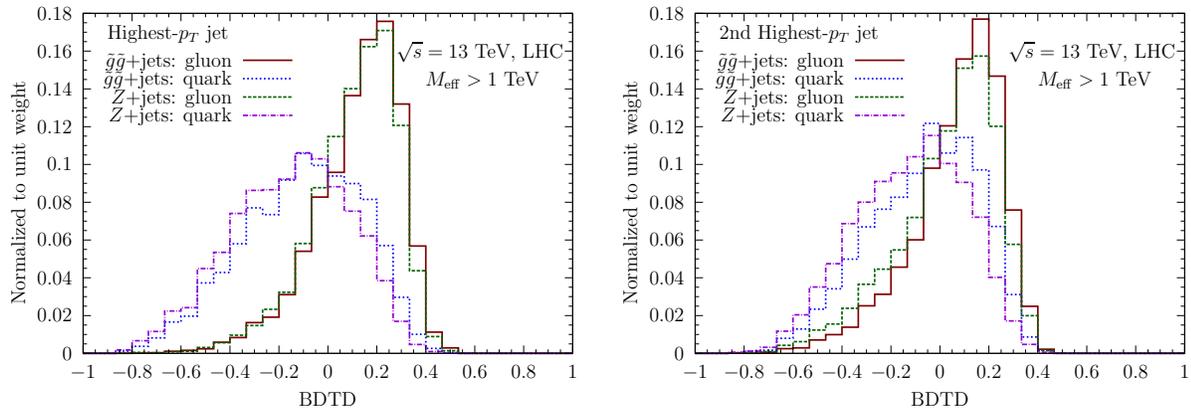

\begin{center}
\centerline{\resizebox{8cm}{!}{\input{qgj1.tex}} 
\resizebox{8cm}{!}{\input{qgj2.tex}} }
\caption{\small \sl Normalized (to unit weight) distribution of the BDTD variable for the $\gl \gl$ signal (gluon: solid red, quark: blue dotted) and the $Z$-background (gluon: green dashed, quark: violet dot-dashed): for the highest-$p_T$ jet (left) and for the second highest-$p_T$ jet (right), after the jet-$p_T$, $\met$ and the $M_{\rm eff}> 1$ TeV cuts, for $13$ TeV LHC. The quark and gluon tags are obtained from Monte Carlo truth level information. }
\label{fig:BDT}
\end{center}
\end{figure}

The training of the classifier was performed with $Z+q-$jet and $Z+g-$jet samples and we generated the above Monte Carlo samples uniformly distributed in jet-$p_T$. We define 10 different categories by the jet $p_T$'s, with $N_{\rm Tracks}$ and $w_j$ as the input variables for the training. In Fig.~\ref{fig:BDT} we show the normalized (to unit weight) distribution of the decorrelated BDT variable (BDTD) for the $\gl \gl+$jets signal and the $Z+$jets background events after the jet-$p_T$, $\met$ and $M_{\rm eff}>1$ TeV cuts (as described in Table~\ref{xsec}). 
%%%%%%%%%%%%%%%%%%%%%%%%%%%%%%%%%%%%%%%%%%
%%%%%%%%%%%%%%%%%%%%%%%%%%%%%%%%%%%%%%%%%
For the identification of the jets as quark or gluon initiated ones, we have used the Monte Carlo truth level information. The distributions are shown separately for both the highest $p_T$ (left panel) and 2nd highest $p_T$ (right panel) jets. This figure demonstrates that the BDTD variable can effectively discriminate between a quark jet and a gluon jet, and therefore is a validation of the proper training of the classifier. Furthermore, the discrimination capability is seen to be similar for the signal and background processes. In order to estimate the actual quark and gluon jet fractions in the signal and the $Z$ background after the cuts described above, we again appeal to the Monte Carlo truth level information, and the results are shown in Table~\ref{tab:qg}. 

%%%%%%%%%%%%%%%%%%%%%%%%%%%%%%%%%%%%%%%%%%%%%%%%%%%%%%%%%%%%%%%%%%%%%%%%%%%%%%%%%
\begin{table}[htb!]
\centering
\begin{tabular}{|c|c c|c c|}
\hline
Process & \multicolumn{2}{c|}{Highest-$p_T$ jet}   & \multicolumn{2}{c|}{2nd highest-$p_T$ jet}  \\
              & $f_g$ & $f_g^{~{\rm BDTD}>0.15}$ &  $f_g$ & $f_g^{~{\rm BDTD}>0.15}$  \\
\hline
$\gl \gl +$jets& 0.46       &0.73 &  0.81                & 0.90   \\

$Z+$jets &   0.35           &0.70 &      0.65             & 0.84  \\

\hline

\end{tabular}
\caption{\small \sl Gluon fraction ($f_g$) at MC truth level before and after the BDTD cut, for the highest and 2nd highest-$p_T$ jets in $\gl \gl +$jets and $Z+$jets processes. All events are selected after passing the jet-$p_T$, $\met$ and $M_{\rm eff}>1$ TeV cuts, at the 13 TeV LHC.}
\label{tab:qg}
\end{table}
%%%%%%%%%%%%%%%%%%%%%%%%%%%%%%%%%%%%%%%%%%%%%%%%%%%%%%%%%%%%%%%%%%%%%%%%%%%%%%%%%%%%%
For both the high $p_T$ jets considered we can see that the signal has a higher gluon fraction ($f_g$) compared to the $Z+$jets background (the quark fraction is $f_q=1-f_g$). Moreover, $f_g$ is seen to be higher for the 2nd highest $p_T$ jet. Based on Fig.~\ref{fig:BDT} we use a cut on the BDTD variable for both the jets, ${\rm BDTD}>0.15$ to enhance the $S/B$ ratio. As expected, the gluon jet fraction $f_g$ is enhanced significantly after this cut, as seen from the $f_g^{~{\rm BDTD}>0.15}$ columns in Table~\ref{tab:qg} (the enhancement is more pronounced for the highest $p_T$ jet as the separation is better, see Fig.~\ref{fig:BDT}). The efficiency of this cut on BDTD is shown in Table~\ref{tab:qgeff}, where, $\epsilon_{j_1}$, $\epsilon_{j_2}$ and $\epsilon_{\rm Total}$  represent the efficiency of the ${\rm BDTD}>0.15$ cut on the highest $p_T$ jet, the 2nd highest $p_T$ jet and the combined efficiency for a cut on both the jets respectively. Due to the higher fraction of gluon jets in the signal, the efficiencies are higher for the $\gl \gl+$jets process compared to $Z+$jets. The cross-section for $\gl \gl+\geq 2-$jets (signal Point-A) and $Z+\geq 2-$jets after the BDTD cut at 13 TeV LHC are shown in Table~\ref{tab:qgeff} as well. Comparing these to the numbers after $M_{\rm eff}>1$ TeV in Table~\ref{xsec}, we see that there is an improvement in the $S/B$ ratio from $0.1$ to $0.22$. Therefore, utilizing the quark and gluon jet discrimination based on a BDT analysis can help us further improve the search for degenerate gauginos at the LHC. It should be mentioned here that a recent study by the ATLAS collaboration on light quark and gluon jet discrimination with $7$ TeV LHC data~\cite{Atlas-qg} finds some differences between the tagging efficiencies found in the data and the predictions of the {\tt PYTHIA6} or {\tt HERWIG++}~\cite{Herwig} Monte Carlo (MC) generators. However, the systematic uncertainty in the jet-tagger performance is still quite large, and future improvements in the analysis may clarify the situation better. In order to estimate the uncertainty in the MC predictions and how it affects the expected improvement factors in SUSY search, a detailed comparison between the results from the two MC generators above is necessary, and we shall report it in a future study. 
%%%%%%%%%%%%%%%%%%%%%%%%%%%%%%%%%%%%%%%%%%%%%%%%%%%%%%%%%%%%%%%%%%%%%%%%%%%%%%%%%
\begin{table}[htb!]
\centering
\begin{tabular}{|c|c c|c c|}
\hline
Process & $\epsilon_{j_1}$ & $\epsilon_{j_2}$  & $\epsilon_{\rm Total}$ & $\sigma_{{\rm BDTD}>0.15}$ \\
\hline
$\gl \gl +$jets& 0.32       & 0.34             &0.11                  &7.86 fb    \\

$Z+$jets &     0.23         & 0.25             &0.05                   &36.49 fb  \\

\hline

\end{tabular}
\caption{\small \sl Efficiency of the ${\rm BDTD}>0.15$ cut on the highest-$p_T$ jet ($\epsilon_{j_1}$) and the 2nd highest $p_T$ jet ($\epsilon_{j_2}$), for $\gl \gl +$jets and $Z+$jets at the 13 TeV LHC . The combined efficiency of both the cuts ($\epsilon_{\rm Total}$), as well as the total cross-section after the BDTD cut are also shown (the signal cross-section is for Point-A). The BDTD cuts were applied on events passing the jet-$p_T$, $\met$ and $M_{\rm eff}>1$ TeV cuts.}
\label{tab:qgeff}
\end{table}
%%%%%%%%%%%%%%%%%%%%%%%%%%%%%%%%%%%%%%%%%%%%%%%%%%%%%%%%%%%%%%%%%%%%%%%%%%%%%%%%

The primary difficulties in combining the BDTD cut with the cuts found in the previous sub-sections (especially $\Delta \eta_{j_1 j_2}$) are twofold. First of all, even though we significantly improve the $S/B$ ratio using both set of cuts, the total signal cross-section drops considerably in both cases. Combining them will result in a further reduction of the signal events giving rise to poor signal statistics. Secondly, one of the variables used by us for quark-gluon discrimination is the number of charged tracks inside the jet cone, which can be evaluated only if $|\eta_j|<2.5$, as determined by the tracker coverage in the LHC detectors. On the otherhand, the $|\Delta \eta_{j_1 j_2}|$ and $\Delta \phi_{j_1 j_2}$ cuts are designed for jets widely separated in rapidity in a VBF-type event topology, which are very often in the forward region, and hence outside the coverage of the tracker. It will be interesting to study whether an optimization using all the relevant cuts is possible, which, however, is beyond the scope of the present work.

\subsection{Discovery and exclusion reach in gluino mass}
Having discussed the effects of various sets of kinematic cuts, we are now in a position to evaluate the discovery or exclusion reach in gluino mass at the 13 TeV LHC. In this connection, it is important to consider the systematic uncertainty ($\Delta B$) in the SM background predictions. Since we are unable to make a quantitative estimate of this uncertainty, which will be carried out in future by the experimental collaborations, we shall present our conclusions assuming it to be in the range from a negligible number to a maximum of $20\%$. We also do not include the effect of the BDTD discriminant in this sub-section as our study of this variable was of an exploratory nature, and the associated systematics  can be different and even higher than that of the standard cut-based analysis. For Point-A with $(M_{\gl},M_{\neu})=(800,780)$ GeV, we find that for $\Delta B=0$ and $0.1$, a $2\sigma$ exclusion is possible with around $10 {~\rm to ~}20 {~\rm fb}^{-1}$ luminosity, after Cuts A and B respectively. With $\Delta B=0.2$ and $50 {~\rm fb}^{-1}$ of data, a $1.8\sigma$ exclusion can be reached, while the significance asymptotically reaches $2\sigma$ only after a large luminosity of $\sim 225 {~\rm fb}^{-1}$ is gathered. A $5 \sigma$ discovery can be achieved for this point if the systematic uncertainty can be reduced to $\mathcal{O}(7\%)$ level, with $300 {~\rm fb}^{-1}$ of data and using Cut C. For Point-B with $(M_{\gl},M_{\neu})=(1000,980)$ GeV, we can only achieve a $2 \sigma$ exclusion within $300 {~\rm fb}^{-1}$ if $\Delta B=0$ (with Cut C). The discovery or exclusion prospects using our methodology is very similar to that obtained by other authors~\cite{Biplob,Wacker} employing different techniques. It is conceivable that an optimized combination of the different discriminating variables would help us achieve a better combined reach in the compressed SUSY parameter space. 

\section{Summary}
A compressed gaugino spectrum can be realized in certain SUSY breaking scenarios, and the LHC bounds on the gluino mass are considerably weaker in such a case. As an example, we briefly review a well-motivated SUSY axion model which can lead to a deformation of the anomaly mediation prediction for gaugino masses and give rise to nearly degenerate gluinos and winos. The primary purpose of this study is to explore  topology-based search strategies for a compressed gluino-neutralino system at the 13 TeV LHC, which can be used in combination with the standard $\met$ and $M_{\rm eff}$ variables. We study the prospects of using rapidity and azimuthal angle correlations between the highest $p_T$ ISR jets. These correlations between the tagged jets can be utilized by focusing on a VBF-type signal topology, with at least two jets, no leptons and $\met$ in the final state. The rapidity separation between the tagged jets is found to be an important variable, and a cut of $|\Delta \eta_{j_1 j_2}|>3.5$   enhances the $S/B$ ratio, and consequently the reach in gluino mass considerably. In particular, for the $t \bar{t}+$jets background, the $\Delta \eta_{j_1 j_2}$ distribution is found to be sharply peaked at lower values, and falls off significantly by $\Delta \eta_{j_1 j_2} \sim 2$. In the signal process of $\gl \gl +$jets, there is a relative excess over the $V+$jets ($V=Z,W$) background for $\Delta \eta_{j_1 j_2}>2.5$. Since higher values of $\met$ or $M_{\rm eff}$ are difficult to obtain in a compressed scenario, the $\Delta \eta_{j_1 j_2}$  cut is found to be more tailored to the signal topology.

After a cut on the $\Delta \eta_{j_1 j_2}$ variable, we find a distinct correlation in the distribution of the azimuthal angle difference between the tagged jets ($\Delta \phi_{j_1 j_2}$). The $\Delta \phi_{j_1 j_2}$ distribution for $\gl \gl+$jets peaks at around $\pi/2$, falling off by $\pi$. The distribution for $Z+$jets, on the otherhand, is rather flat and has a small rise from $0$ to $\pi$. The two $\Delta \phi_{j_1 j_2}$ distributions cross-over at around  $\sim \pi/2$, and therefore, a cut on this variable, $\Delta \phi_{j_1 j_2}<\pi/2$, helps enhance the $S/B$ ratio further. The $\Delta \phi_{j_1 j_2}$ variable is not only helpful for background reduction, it will be interesting to study such azimuthal angle correlations in the aftermath of an actual discovery. As is well-known, the $\Delta \phi_{j_1 j_2}$ distribution in a VBF topology carries the information of the spin and CP quantum numbers of the centrally produced heavy particles, in this case of the gluinos.

After studying various combinations of $M_{\rm eff},\Delta \eta_{j_1 j_2}$ and $\Delta \phi_{j_1 j_2}$ cuts, we conclude that an $800$ GeV gluino (with $M_{\neu}=780$ GeV) can be excluded at $95\%$ C.L. with an integrated luminosity of $20 {~\rm fb}^{-1}$, including the effect of a  systematic uncertainty of $10\%$ on the background, while for a larger systematic uncertainty of $\mathcal{O}(20\%)$, more luminosity ($\sim 225 {~\rm fb}^{-1}$) is necessary. For heavier masses, a $1$ TeV gluino can be excluded at $2\sigma$ with an integrated luminosity of $300 {~\rm fb}^{-1}$ considering only statistical uncertainties.

We further explored the application of quark and gluon jet tagging methods, to utilize the fact that the $\gl \gl+$jets signal has a higher fraction of ISR gluon jets compared to the primary background of $Z+$jets. We used the number of charged tracks inside the jet radius and the width of the jet as the discriminating variables. In order to deal with the non-linear boundary in the plane of these two variables that separate the signal and background regions, we employed a boosted decision tree algorithm using the {\tt TMVA Toolkit} within the {\tt ROOT} analysis framework. It is observed that a cut on the BDTD variable (${\rm BDTD}>0.15$) can enhance the $S/B$ ratio by around a factor of $2$, where the BDTD cut is applied after the jet-$p_T$, $\met$ and $M_{\rm eff}>1$ TeV cuts. It is therefore promising to employ such quark-gluon tagging algorithms in searching for compressed gauginos. The primary difficulty faced by us while trying to combine this technique with the kinematic correlations is the large reduction in signal statistics in both methodologies. An optimization between the two might be a possibility, and we expect that both the kinematic correlations and quark-gluon jet tagging methods discussed in this study will be further investigated by the ATLAS and CMS collaborations to boost the degenerate gluino-neutralino search prospects at the 13 TeV LHC.

%%%%%%%%%%%%%%%%%%%%%%%%%%%
%%%%% Acknowledgments %%%%%
%%%%%%%%%%%%%%%%%%%%%%%%%%%
\section*{Acknowledgments}
We thank Bryan Webber for clarifying aspects of parton shower and matrix element matching. SM thanks Biplob Bhattacherjee for useful discussions on Ref.~\cite{Biplob}. This work is supported by the Grant-in-Aid for Scientific Research from the Ministry of Education, Science, Sports, and Culture (MEXT), Japan (No. 23104006 for M.M. Nojiri, and No. 22244021 for T.T.Yanagida), and also by the World Premier International Research Center Initiative (WPI Initiative), MEXT, Japan.

%%%%%%%%%%%%%%%%%%%%%%
%%%%% References %%%%%
%%%%%%%%%%%%%%%%%%%%%%

\end{document}